\documentclass[a4paper,fleqn]{cas-dc}

\usepackage[numbers]{natbib}

\usepackage{lineno,hyperref}
\modulolinenumbers[5]

\usepackage[nomargin,inline,marginclue,draft]{fixme}

\usepackage{booktabs, multirow} 
\usepackage{soul}
\usepackage{changepage,threeparttable} 

\usepackage[T1]{fontenc}
\usepackage{amsmath,amsthm,amssymb,latexsym}
\usepackage{ragged2e}
\usepackage{caption}
\usepackage{subcaption}
\usepackage{tikz} 
\usepackage{datetime}
\usepackage{epsfig}
\usepackage{graphicx}
\usepackage[utf8x]{inputenc}
\usepackage{slashbox} 
\usepackage{adjustbox} 
\usepackage{cuted} 
\usepackage{float} 
\restylefloat{table}

\DeclareGraphicsExtensions{.bmp,.png,.pdf,.jpeg,.eps}

\graphicspath{{./Figs/}}

\begin{document}
\let\WriteBookmarks\relax
\def\floatpagepagefraction{1}
\def\textpagefraction{.001}

\shorttitle{Critical temperature of 1D Ising chains with long-range interactions} 

\shortauthors{J.G. Mart\'inez-Herrera, et al.}

\title [mode = title]{Critical temperature of one-dimensional Ising model with long-range interaction revisited}                  

\author[1]{J. G. Mart\'inez-Herrera}[type=editor,
orcid=0000-0001-6992-9262]
\ead{mhjguillermo@gmail.com}


\affiliation[1]{
	organization={Posgrado en Ciencias F\'isicas, Universidad Nacional Aut\'onoma de M\'exico},
	addressline={Apdo. Postal 20-364, 01000 Ciudad de M\'exico, M\'exico}
}

\author[2]{O. A. Rodr\'iguez-L\'opez}[type=editor,
orcid=0000-0002-3635-9248]
\ead{oarodriguez.mx@gmail.com}


\author[2]{M. A. Sol\'is}[
]
\ead{masolis@fisica.unam.mx}


\affiliation[2]{organization={Instituto de F\'isicas, Universidad Nacional Aut\'onoma de M\'exico},
	addressline={Apdo.
		Postal 20-364, 01000 Ciudad
		de M\'exico, M\'exico}
}

\begin{abstract}
We present a generalized expression for the transfer matrix of finite and infinite one-dimensional spin chains within a magnetic field with spin pair interaction $J/r^p$, where $r \in \{ 1,2,3,\ldots,n_v \}$ is the distance between two spins, $n_v$ is the number of nearest neighbors reached by the interaction, and $1 \leq p \leq 2$. Using this transfer matrix, we calculate the partition function, the Helmholtz free energy, and the specific heat for both finite and infinite ferromagnetic 1D Ising models within a zero external magnetic field. We focus on the temperature $T_{\text{max}}$ where the specific heat reaches its maximum, needed to compute $J/(k_B T_{\text{max}})$ numerically for every value of $n_v \in \{ 1,2,3,\ldots, 30\}$, which we interpolate and then extrapolate up to the critical temperature as a function of $p$, by using a novel inter-extrapolation function. We use two different procedures to reach the infinite spin chain with an infinite interaction range: increasing the chain size as well as the interaction range by the same amount and increasing the interaction range for the infinite chain. As we expected, both extrapolations lead to the same critical temperature within their uncertainties by two different concurrent curves. Critical temperatures fall within the upper and lower bounds reported in the literature, showing a better coincidence with many existing approximations for $p$ close to 1 than the $p$ values near 2.
It is worth mentioning that the well-known cases for nearest (original Ising model) and next-nearest neighbor interactions are recovered doing $n_v = 1$ and $n_v = 2$, respectively.
\end{abstract}


\maketitle


\section{Introduction}
 Since its solution in 1925, the Ising model has been studied in many facets to understand systems with interactions, paying special attention to phase transitions. Although the original 1D Ising model does not present a spontaneous magnetization at non-zero temperature, it has been proved~\cite{Dyson-1968,Frohlich-1982} that the Ising model with long-range interactions shows a phase transition at finite temperature. The first demonstration of a phase transition existence in the ferromagnetic long-range interaction Ising chain with an exchange interaction $\mathcal{J}(r) = J/r^p$ was given by Dyson~\cite{Dyson-1968} for $1 < p < 2$, as well as the non existence of phase transition for $p > 2$~\cite{Dyson-1968-January,Dyson-1971} but keeping the case $p=2$ in doubts. The distance $r$ between interacting spins is given in units of the nearest neighbors and $J$ is a constant energy to which $\mathcal{J}(r)$ reduces when $p \to 0$. After some years, in 1982, Fröhlich and Spencer~\cite{Frohlich-1982} proved the existence of a phase transition for $p=2$. Since the original demonstration of phase transition existence by Dyson, who did not provide a critical temperature, many approximations have been reported to give the critical temperatures for $1 \leq p \leq 2$ using different methods: variational~\cite{Wragg-Gehring-1990}, coherent anomaly~\cite{Monroe-1990}, $\zeta$-function~\cite{Mainieri-1992}, finite-range~\cite{Gluman-Uzelac-1989}, series expansion~\cite{Nagle-Bonner-1969}, re-normalization group~\cite{Anderson-Yuval-1971,Cannas-1995}, Monte Carlo~\cite{Bhattacharjee-1981}, Bethe Lattice approximation~\cite{Monroe-1992}, cluster approach~\cite{Doman-1981}, among others~\cite{Matvienko-1985,Vigfusson-1986,Uzelac-1988,Pires-1996,Krech-2000,Luijten-1997,Luijten-2001}. 

Moreover, the value of $p$ determines the type of transition on the system~\cite{Luijten-1997}. For $p\in (1,1.5)$, the system presents a phase transition of the mean-field type, i.e., the specific heat presents a jump at the transition~\cite{Krech-2000}, and the critical exponents are temperature independent as obtained by a mean-field approach~\cite{Aizenman-Fernandez-1988}. For $p \in (1.5,2)$, the phase transition is non-mean-field type, and the specific heat presents a sharp peak~\cite{Krech-2000}. For $p=2$, the system exhibits a jump in the magnetization~\cite{Aizenman-1988,Dyson-1971}, which is known as the Thouless effect~\cite{Thouless-1969} and the spin-spin correlation function presents a power-law decay with a temperature-dependent exponent, which is related to the Kosterlitz-Thouless transition~\cite{Fukui-2009}.

Although a phase transition can be identified by a discontinuity or divergence in the specific heat or the magnetic susceptibility, we focus on the specific heat since the magnetic susceptibility for spins systems has a very abrupt change at $h=0$, making it difficult its calculate. In addition, specific heat is a recurrently measured quantity in both physical and computational experiments to study phase transitions~\cite{Chen-Gasparini,Ferdinand-1969,Merdan-2004,Bexter-1979,Bauer-2013}. 

Despite a large amount of work to obtain the critical temperature of the long-range ferromagnetic 1D Ising model, there is still no \textit{exact} analytical expression for its critical temperature.
Our goal in this work is to give two new extrapolated critical temperature curves as functions of $p$ from the extrapolation of the temperatures at which the specific heat has a maximum. For this, we propose a novel inter-extrapolation function inspired by the Hurwitz zeta function. Inter-extrapolations are carried out in two ways: i) from an infinite system with finite range interactions to a system of infinite range interactions, and ii) from a finite system increasing the size and the range of interactions. 

The rest of the article is organized as follows: In Sec. II, we generalize the Kassan-Ogly transfer matrix~\cite{Kassan-2001} for the ferromagnetic long-range Ising model to include any number $n_v$ of nearest neighbors, which we use to obtain the corresponding partition function, the Helmholtz free energy and the specific heat expression used to calculate the temperature where the specific heat shows a maximum, which we called $T_{\text{max}}$. In Sec. III, we use the set of $T_{\text{max}}$ obtained for each $n_v \in \{ 1,2,3,\ldots, 30\}$ value to calculate the critical temperature as an extrapolation of $J/(k_BT_{\text{max}})$ vs. $1/n_v$ as $n_v \rightarrow \infty$ for each $p \in \{1,1.05,1.1,1.2,1.3,\ldots,2\}$. In Sec. IV, we give our conclusions. We summarize our numerical calculations in Tables~\ref{tab:all_Tc},~\ref{tab:finite_all_Tc},~\ref{Tab:Hurwitz_params}, and~\ref{Tab:Hurwitz_Tc} given in Appendix~\ref{Appendix:Tables}, and we give additional details of the calculation code and the comparisons of our inter-extrapolation functions in Appendix~\ref{Appendix:Program specification}.

\section{Ferromagnetic long-range Ising Model}

The Hamiltonian of the 1D ferromagnetic Ising model with $n_v$-neighbors interactions is
\begin{equation}\label{SystemHamiltonian}
	H=-\sum_{i=1}^{N} \sum_{j=1}^{n_v}J_{i,j}\sigma_{i}\sigma_{i+j} - h\sum_{i=1}^N \sigma_{i},
\end{equation}
where $J_{i,j}$ is the $i,j$-neighbor interaction given by
\begin{equation}\label{SpinInteractions}
	J_{i,j} = \left\{ 
	\begin{array}{ll}
		J \frac{1}{|i-j|^p} & \text{if} \, |i-j|\leq n_v \\
		0                   & \text{in other case}
	\end{array}
	\right.,
\end{equation}
$N$ is the number of spins, and $h$ is the energy interaction with the external magnetic field.
In the case of periodic boundary conditions, $\sigma_{N+k}=\sigma_k$.

The partition function for this system is given by
\begin{equation}
	Z=\sum_{{\{\sigma_i =\pm1\}}_{i=1}^N} \exp[-\beta H]
\end{equation}
where $\beta=1/(k_BT)$.

In this case, we can define the transfer matrix $W$, whose elements are given by
\begin{equation}\label{GeneralMatrixElements}
	\begin{split}
		\langle \sigma_{k}, & \sigma_{k+1},\cdots,\sigma_{k+n_v-1}|  W|\sigma_{k+1},\sigma_{k+2},\cdots,\sigma_{k+n_v}\rangle = 
		\\ & \exp\left[\beta \sum_{j=1}^{n_v} J_{i,j}\sigma_{k}\sigma_{k+j} + \beta h\sigma_{k}\right], \, \, \, k=1,2,\ldots,N,
	\end{split}
\end{equation}
where the transfer matrix elements with incompatibles spin projections (two different values for the same $\sigma_k$) are zero. This transfer matrix (\ref{GeneralMatrixElements}) is a generalization for any $n_v$ value of the Kramers-Wannier one~\cite{Kassan-2001}.

For example, the matrix element $\langle \sigma_1=1,\sigma_2=1|W|\sigma_2=-1,\sigma_3=1\rangle = 0$ because the $\sigma_{2}$ can not take values $1$ and $-1$ at the same time. Thus, the transfer matrix for $n_v=2$ is 

%
\begin{strip}
	\begin{equation}
		\resizebox{0.9\hsize}{!}{$
			\begin{split}
				& W = \begin{pmatrix}
					\langle 1,1|W|1,1 \rangle   & \langle 1,1|W|1,-1 \rangle   & \langle 1,1|W|-1,1 \rangle   & \langle 1,1|W|-1,-1 \rangle   \\
					\langle 1,-1|W|1,1 \rangle  & \langle 1,-1|W|1,-1 \rangle  & \langle 1,-1|W|-1,1 \rangle  & \langle 1,-1|W|-1,-1 \rangle  \\
					\langle -1,1|W|1,1 \rangle  & \langle -1,1|W|1,-1 \rangle  & \langle -1,1|W|-1,1 \rangle  & \langle -1,1|W|-1,-1 \rangle  \\
					\langle -1,-1|W|1,1 \rangle & \langle -1,-1|W|1,-1 \rangle & \langle -1,-1|W|-1,1 \rangle & \langle -1,-1|W|-1,-1 \rangle
				\end{pmatrix}
				\\ & 
				\\ & = \begin{pmatrix}
					\exp\left[\beta(J+\frac{J}{2^p}+h)\right]  & \exp\left[\beta(J-\frac{J}{2^p}+h)\right]  & 0                               & 0                               \\
					0                               & 0                               & \exp\left[\beta(-J+\frac{J}{2^p}+h)\right] & \exp\left[\beta(-J-\frac{J}{2^p}+h)\right] \\
					\exp\left[\beta(-J-\frac{J}{2^p}-h)\right] & \exp\left[\beta(-J+\frac{J}{2^p}-h)\right] & 0                               & 0
					\\
					0                               & 0                               & \exp\left[\beta(J-\frac{J}{2^p}-h)\right]  & \exp\left[\beta(J+\frac{J}{2^p}-h)\right]
				\end{pmatrix}.
			\end{split}
			$}
	\end{equation}
	However, in a finite chain with an interaction range $n_v$ equal to the number of the spins $N=2$, the transfer matrix elements are $\langle \sigma_{1},\sigma_{2} | W | \sigma_{2},\sigma_{1} \rangle$. Then, there are only four non-zero elements, i.e., 
	\begin{equation}
		\resizebox{0.9\hsize}{!}{$
			\begin{split}
				& W = \begin{pmatrix}
					\exp\left[\beta(J+\frac{J}{2^p}+h)\right]  & 
					0                               & 
					0                               & 0                               \\
					0                               & 0                               & \exp\left[\beta(-J+\frac{J}{2^p}+h)\right] &
					0 \\
					0 								& 
					\exp\left[\beta(-J+\frac{J}{2^p}-h)\right] & 0                               & 
					0
					\\
					0                               & 0                               & 
					0  								& 
					\exp\left[\beta(J+\frac{J}{2^p}-h)\right]
				\end{pmatrix}.
			\end{split}
			$}
	\end{equation}
\end{strip}

With this transfer matrix definition, the partition function becomes
\\
\begin{equation}\label{FinitePartitionFuntion}
	\resizebox{0.9\hsize}{!}{$
		\begin{split}
			Z & = \sum_{{\{\sigma_i =\pm1\}}_{i=1}^N} \prod_{k=1}^N\langle\sigma_{k},\sigma_{k+1},\cdots,\sigma_{k+n_v-1}|W|\sigma_{k+1},\sigma_{k+2},\cdots,\sigma_{k+n_v}\rangle
			\\ & = Tr(W^N) = \sum_{i=1}^{n_e} \lambda_{i}^N = \lambda_{\text{max}}^N \sum_{i=1}^{n_e} \left( \frac{\lambda_{i}}{\lambda_{\text{max}}} \right)^N,
		\end{split}
		$}
\end{equation}
where $n_e=2^{n_v}$ is the number of eigenvalues of the matrix $W$, $\{\lambda_i\}_{i=1}^{n_e}$ are the eigenvalues of the transfer matrix $W$, and $\lambda_{\text{max}}$ is the greatest eigenvalue. 

Using the relation for the Helmholtz free energy $F = -k_B T \ln Z$, we obtain for finite chains
\begin{equation}\label{HelmholtzFreeEnergy}
	F = - Nk_BT \, \ln\left[\lambda_{\text{max}}\right] - k_BT \, \ln\left[\sum_{i=1}^{n_e}(\lambda_{i}/\lambda_{\text{max}})^N\right],
\end{equation}
from where we get the magnetization $M = -\left( \partial F/\partial h \right)_{T,N}$ as
\begin{equation}
	\begin{split}
		M/N = & \frac{k_BT}{\lambda_{\text{max}}}\left(\frac{\partial \lambda_{\text{max}}}{\partial h}\right)_{T,N} \\ & - \frac{k_BT}{\sum_{i=1}^{n_e}(\lambda_i/\lambda_{\text{max}})}\left( \sum_{i=1}^{n_e}\frac{\partial (\lambda_i/\lambda_{\text{max}})}{\partial h} \right)_{T,N}.
	\end{split}
\end{equation}
The specific heat comes from  
\begin{equation}\label{SpecificHeat}
	C = - T\left( \frac{\partial^2 F}{\partial T^2} \right)_{h,N}.
\end{equation}

In the thermodynamic limit ($N\rightarrow \infty$), the partition function (\ref{FinitePartitionFuntion}) reduces to
\begin{equation}
	Z=\lambda_{\text{max}}^N,
\end{equation}
then the Helmholtz free energy becomes
\begin{equation}\label{HelmholtzFreeEnergyTL}
	F=-Nk_BT \, \ln[\lambda_{\max}]
\end{equation}
and the magnetization is
\begin{equation}
	\frac{M}{N}=\frac{k_B T}{\lambda_{\text{max}}}\left(\frac{\partial \lambda_{\text{max}}}{\partial h}\right)_{T,N}.
\end{equation}

\section{Phase transition}
The one-dimensional Ising chain with long-range interactions, in the thermodynamic limit, presents an order-disorder phase transition~\cite{Dyson-1971}, which manifests as an anomaly in the specific heat at some temperature.

\begin{figure*}[!ht]
	\begin{subfigure}[b]{1\columnwidth}
		\centering
		\includegraphics[scale=0.5,width=\textwidth]{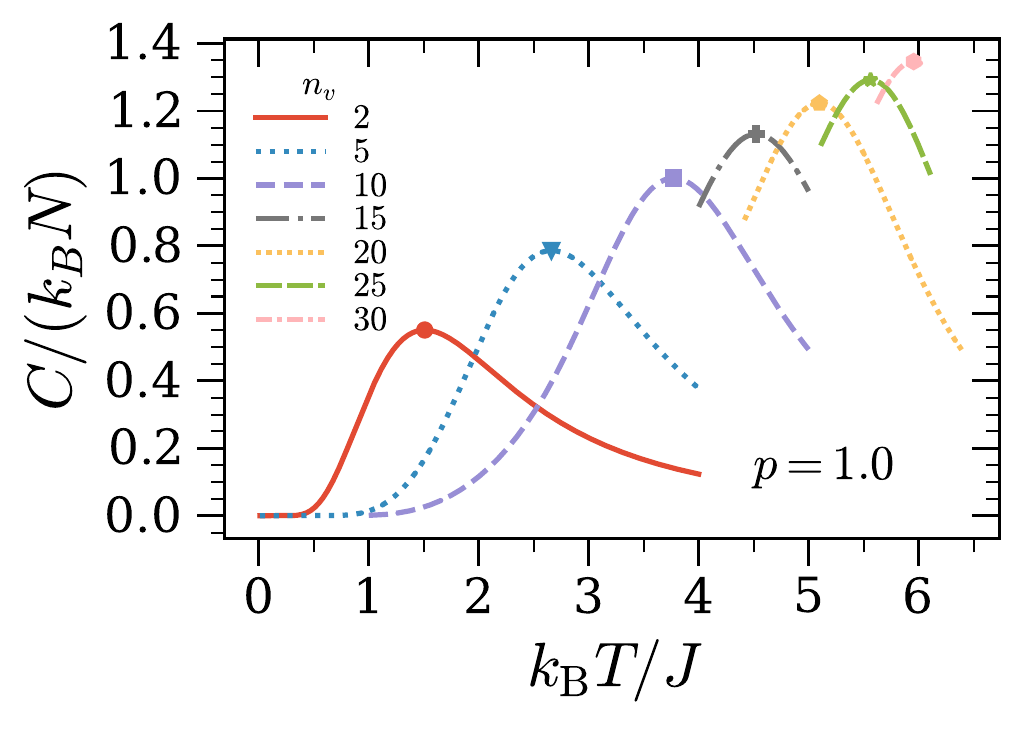}
		\caption{}
		\label{Fig:all_cv_p-1.0}
	\end{subfigure}
	\begin{subfigure}[b]{1\columnwidth}
		\centering
		\includegraphics[scale=0.5,width=\textwidth]{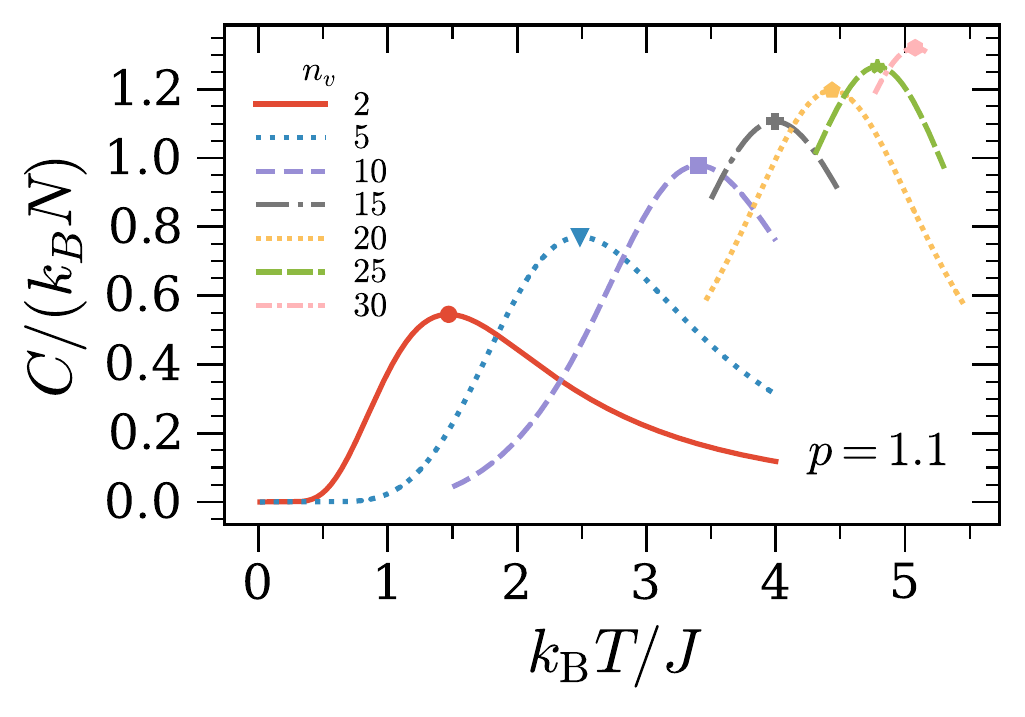}
		\caption{}
		\label{Fig:all_cv_p1.1}
	\end{subfigure}
	\begin{subfigure}[b]{1\columnwidth}
		\centering
		\includegraphics[scale=0.5,width=\textwidth]{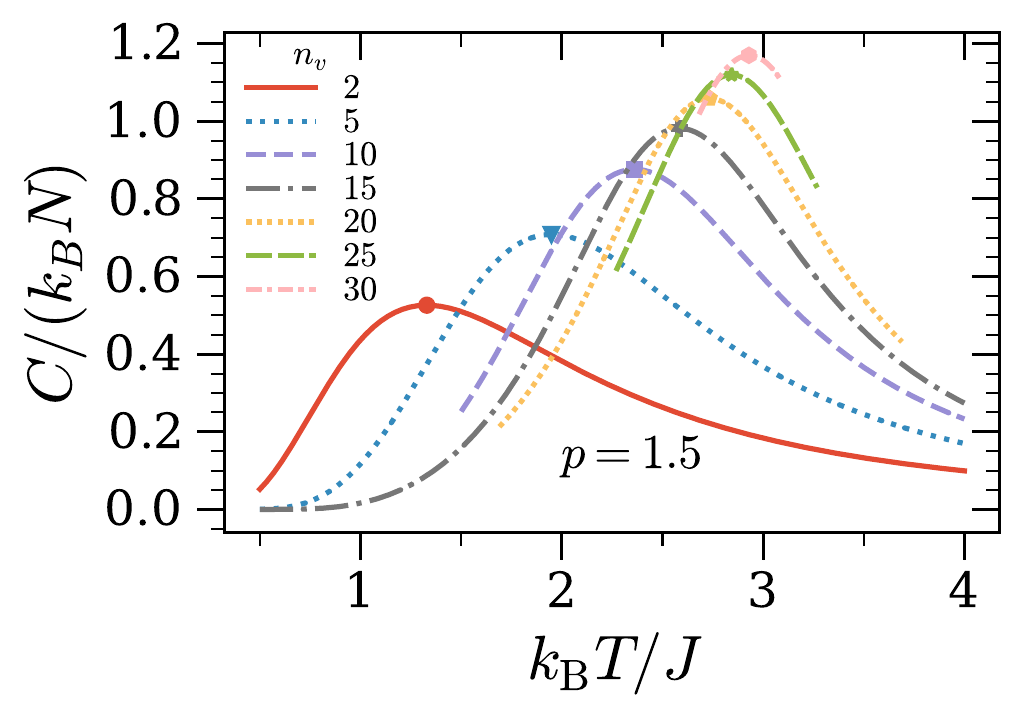}
		\caption{}
		\label{Fig:all_cv_p1.5}
	\end{subfigure}
	\begin{subfigure}[b]{1\columnwidth}
		\centering
		\includegraphics[scale=0.5,width=\textwidth]{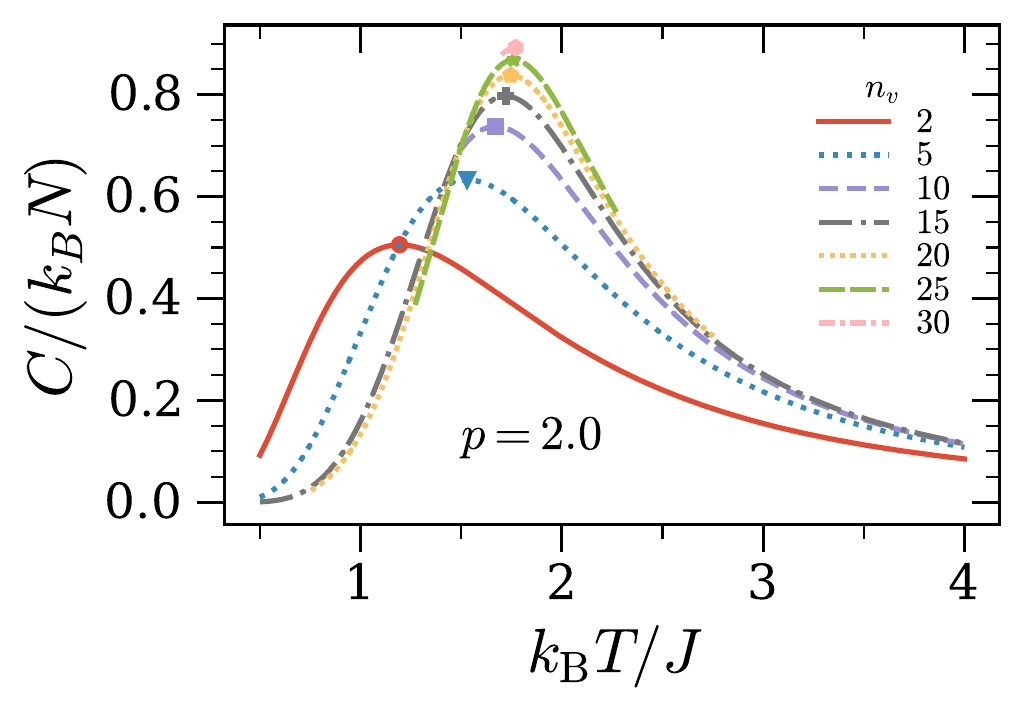}
		\caption{}
		\label{Fig:all_cv_p2.0}
	\end{subfigure}
	\caption{Specific heats as functions of temperature for several values of $n_v$ and a) $p=1.0$, b) $p=1.1$, c) $p=1.5$, and d) $p=2.0$. Here, we can see how curvature around the maximum of the specific heat increases with the $n_v$, and the shape of the curves seems to approach a peak. We also see that the convergences of specific heat maxima increase as $p$ increases.}%
	\label{Fig:Cv_several_p}
\end{figure*}

In Fig.~\ref{Fig:Cv_several_p}, we show the specific heat behavior for infinite chains with different values of $p$ obtained by solving the Eq. (\ref{SpecificHeat}) numerically. We can see that the specific heat maximum increases as the range of interactions $n_v$ increases, approaching a peak. However, these infinite chains with finite $n_v \leq 30$ do not present a phase transition because the specific heat does not exhibit a peak. Also, the chains do not present a pseudo phase transition because the curve shapes are not significantly peaked as observed in a pseudo phase transition (see ~\cite{DeSouza2018}). In Table~\ref{tab:all_Tc} of Appendix~\ref{Appendix:Tables}, we report all $k_BT_{\text{max}}/J$ calculated numerically and their uncertainties. 

In previous work, Dobson~\cite{Dobson-1969} reported a similar calculation of the specific heat for the infinite 1D chain with $n_v \in \{2,3,4,5,6,7\}$ and several values of $p \in \{1.2,1.5,2.0,2.5\}$. Here, we take Dobson's calculation further by increasing $n_v$ until 30. Although the transfer matrix method is easy to implement and flexible to get results for any interaction expression, the time and memory used grow exponentially as the shape of the transfer matrix grows as $2^{n_v} \times 2^{n_v}$. We exploit the centrosymmetric property for $h=0$ to calculate the eigenvalues from a matrix of shape $2^{n_v-1}\times 2^{n_v-1}$ (see Appendix~\ref{Appendix:Program specification}); however, time and memory still grow exponentially, preventing us from obtaining results for larger $n_v$. Going from $n_v=7$ to $n_v=30$ allows us to observe with more accuracy the evolution of critical temperature and propose an inter-extrapolation function to calculate the critical temperature $T_c$.

%
\section{Critical temperature}

From the transfer matrix, the Helmholtz free energy, and the expression (\ref{SpecificHeat}), we compute the specific heats and the $k_BT_{\text{max}}/J$ values numerically with $n_v \in \{2,3,4,\ldots,30\}$ for infinite chains and $n_v \in \{2,3,4,\ldots,12\}$ for finite ones. To obtain the critical temperature for long-range interactions, we extrapolate $J/(k_BT_{\text{max}})$ as a function of $1/n_v$, taking $1/n_v \rightarrow 0$. 

For both finite and infinite chains, we interpolate the $J/(k_B T_{\text{max}})$ values and extrapolate to $J/(k_B T_c)$, with the same inter-extrapolation function, which we propose considering the following facts: for ferromagnetic interactions ($J>0$) where all spins align in the ground state, the energy required to flip anyone is given by 
\begin{equation}\label{EnergyFlipSpin}
	\Delta E = J \sum_{j=1}^{\infty}\frac{1}{j^p}=J\zeta(p),
\end{equation}
where $\zeta(p)$ is the Riemann zeta function. In particular, for $p=1$, we have that the energy needed to disorder the system is $\Delta E = \infty$. Then, there is not a phase transition for any finite temperatures. Moreover, the asymptotic behavior of the critical temperature near $p=1$ reported in Refs.~\cite{Cannas-1995,Hiley-1965} is proportional to Riemann zeta function.

As it is needed an energy proportional to the Riemann zeta function to break the ground state and that asymptotic critical temperature behavior is proportional to the Riemann zeta function, we implement an inter-extrapolation function associated with the Riemann zeta function. Then, we propose the following inter-extrapolation function 
\begin{equation}\label{HurwitzModel}
	\frac{J}{k_BT_{\text{max}}} = \frac{a_1}{h_u(n_v,p,1,a_2)},
\end{equation}
where $a_1$ and $a_2$ are free parameters which we adjust to interpolate all the points $(1/{n_v},J/(k_BT_{\max}))$ for a given $p$ minimizing the dispersion of the points, and
\begin{equation}
	h_u(n_v,p,z,a) \equiv  \sum_{k=0}^{n_v-1} \frac{z^k}{(k+a)^p}.
\end{equation}
To get the critical temperature, we carry out the limit $n_v \rightarrow \infty$ in the expression (\ref{HurwitzModel}) to obtain
\begin{equation}\label{HurwitzExtrapolation}
	\frac{k_BT_{\text{c}}}{J} = \frac{1}{a_1} \hat{\zeta}(p,a_2),
\end{equation}
where 
\begin{equation}
	\hat{\zeta}(p,q) = \sum_{k=0}^{\infty} \frac{1}{(k+q)^p}, \, \, \, q \neq 0,-1,-2,\ldots
\end{equation}
is the Hurwitz zeta function, which is a generalization of the Riemann zeta function satisfying $\zeta(p)=\hat{\zeta}(p,1)$. We observe that this function has the advantage of adjusting quite well the $J/(k_BT_\text{max})$ values for $n_v \geq 3$ and naturally reproduces the asymptotic critical temperature behavior near $p=1$. Moreover, we tried other inter-extrapolation functions as power-laws, conical forms, squared roots, logarithms, and polynomials, but the extrapolated critical temperatures do not respect Monroe's upper and lower bounds~\cite{Monroe1994}, or the least squared method gives less dispersion for the Hurwitz zeta function, see Appendix \ref{Appendix:Program specification}.

\begin{figure}[h!]
	\centering
	\includegraphics[scale=0.85]{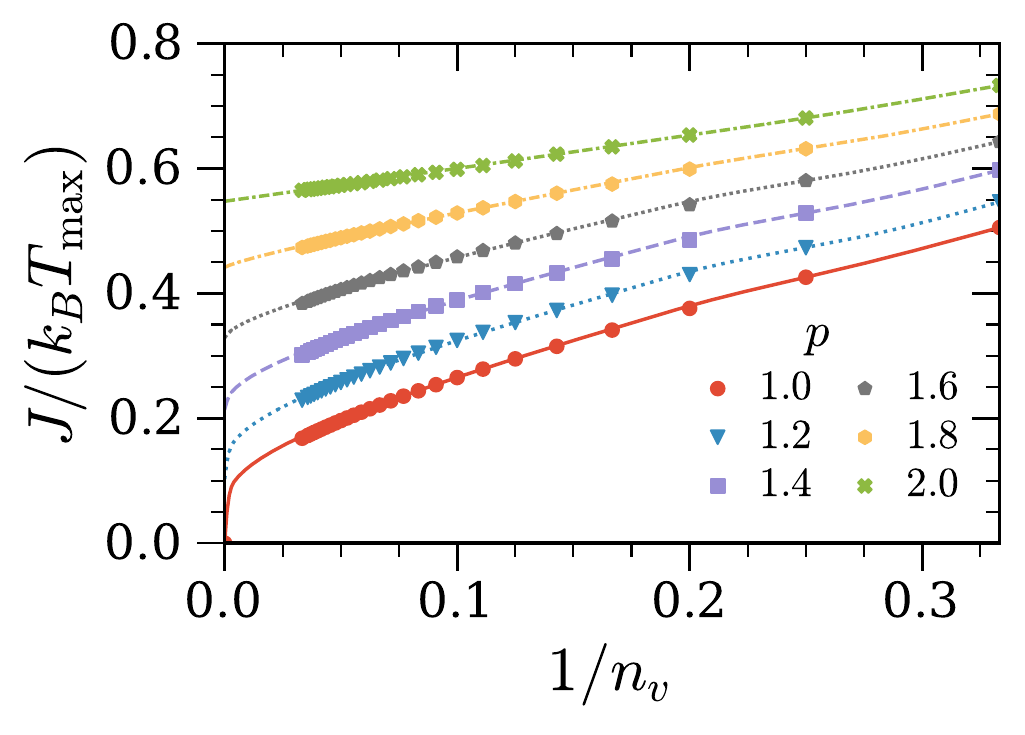}
	\caption{Inter-extrapolation with (\ref{HurwitzModel}) and (\ref{HurwitzExtrapolation}) of $J/(k_BT_{\text{max}})$ infinite chain values as function of $1/n_v$ for several values of $p$.}
	\label{Fig:Extrapol_funct_inv_Tc_vs_inv_nv}
\end{figure}

For infinite chains, in Fig.~\ref{Fig:Extrapol_funct_inv_Tc_vs_inv_nv}, we show several points of $J/(k_BT_{\text{max}})$ as a function of $1/n_v$ for several values of $p$ and the inter-extrapolation function curves. In Table~\ref{Tab:all_temperature_Cv-max} of Appendix~\ref{Appendix:Tables}, we give all $k_BT_{\text{max}}/J$ values with their uncertainties. Note that the $k_BT_{\text{max}}/J$ values at $n_v=1$ must be equal because this case corresponds to the chain with only first neighbors interactions, and, according to Eq. (\ref{SpinInteractions}), the interaction is equal for any value of $p$. This value is $J/(k_BT_{\text{max}}) = 1.19968$, which we can obtain from the analytic specific heat expression (see~\cite{Pathria-2011} p. 479) given by
\begin{equation}
	\frac{C}{Nk_B} = (\beta J)^2 \, \text{sech}^2(\beta J).
\end{equation}
\begin{figure}[h!]
	\centering
	\includegraphics[scale=0.85]{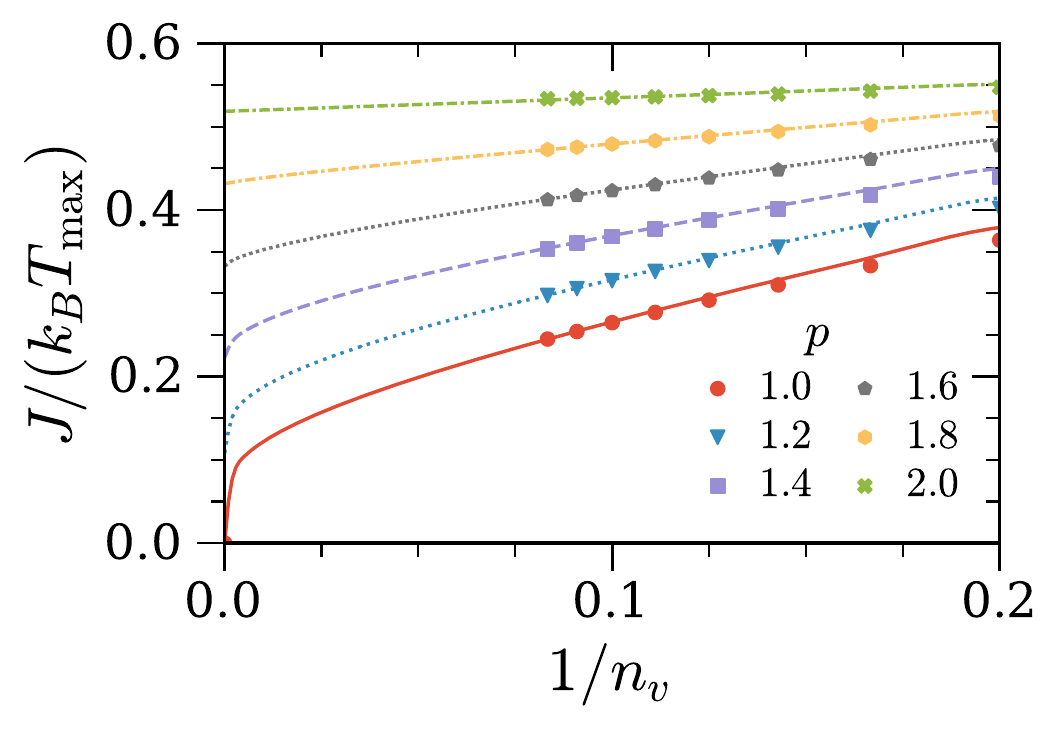}
	\caption{Inter-extrapolation with (\ref{HurwitzModel}) and (\ref{HurwitzExtrapolation}) of $J/(k_BT_{\text{max}})$ finite chain values as function of $1/n_v$ for several values of $p$.}
	\label{Fig:finite_extrapol_funct_inv_Tc_vs_inv_nv}
\end{figure}
For finite chains, in Fig.~\ref{Fig:finite_extrapol_funct_inv_Tc_vs_inv_nv}, we show several points of the $J/(k_BT_{\text{max}})$ as a function of $1/n_v$ and the corresponding inter-extrapolation function curves for $p \in \{1,1.2,1.4,1.6,1.8,2\}$. In Table~\ref{tab:finite_all_Tc} of Appendix~\ref{Appendix:Tables}, we give all $k_BT_{\text{max}}/J$ numerical values with their uncertainties as well as the values of the parameters $a_1$ and $a_2$ in Table~\ref{Tab:Hurwitz_params}. In this case, all values at $n_v=2$ are equal because the specific heat is a $p$-independent function, which can be analytically calculated by
\begin{equation}
	\frac{C}{2k_B} = \frac{1}{2}(2\beta J)^2 \, \text{sech}^2 \left( 2\beta J \right),
\end{equation}
giving $J/(k_BT_{\text{max}}) = 0.599839$ for $n_v=2$.

We obtain the extrapolated inverse critical temperatures from Eq. (\ref{HurwitzExtrapolation}).

\begin{figure}[h!]
	\centering
	\includegraphics[scale=0.85]{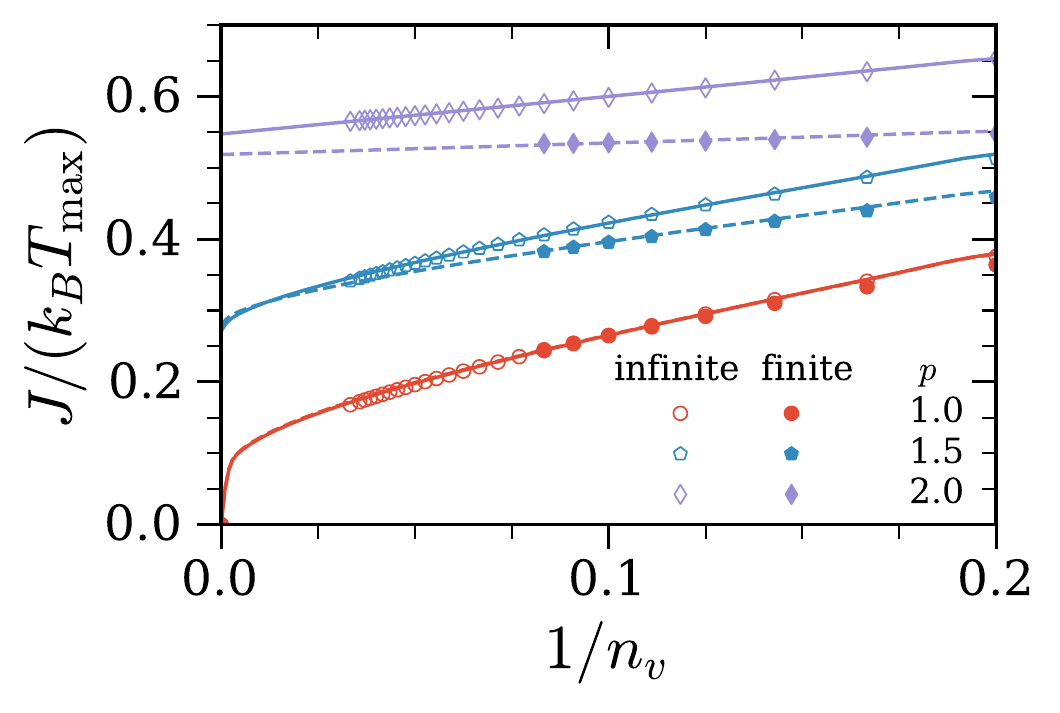}
	\caption{Inter-extrapolations of $J/(k_BT_{\text{max}})$ as functions of $1/n_v$ for several values of $p$ in infinite (solid lines) and finite (dashed lines) chains, coming from (\ref{HurwitzModel}) and (\ref{HurwitzExtrapolation}).}%
	\label{Fig:Tmax_vs_nv_infinite-finiteAllEigs}
\end{figure}

In Fig.~\ref{Fig:Tmax_vs_nv_infinite-finiteAllEigs}, we compare $J/(k_BT_{\text{max}})$ as a function of $1/n_v$ for both infinite and finite chains where inter-extrapolation curves come from Eqs. (\ref{HurwitzModel}) and (\ref{HurwitzExtrapolation}). For $p = 1$, the $J/(k_BT_{\text{max}})$ values for both finite and infinite chains are very close at the same $n_v$ value. Consequently, inter-extrapolated curves are approximately the same. For $p=1.5$, the $J/(k_BT_{\text{max}})$ values present a more significant difference for each $n_v$ value as well as the inter-extrapolated curves for infinite and finite chains, differences which are most remarkable for $p=2.0$. 

Moreover, we observe that the extrapolated curve slope is greater for infinite chains than for finite ones for any $p$, which is more notorious as $p$ increases, i.e., the convergence to $J/(k_BT_c)$ is faster for finite chains than for infinite chains. The difference in convergence speeds decreases as $p$ goes to 1, where the extrapolated curves are indistinguishable. 
\begin{figure}[h!]
	\centering
	\includegraphics[scale=0.85]{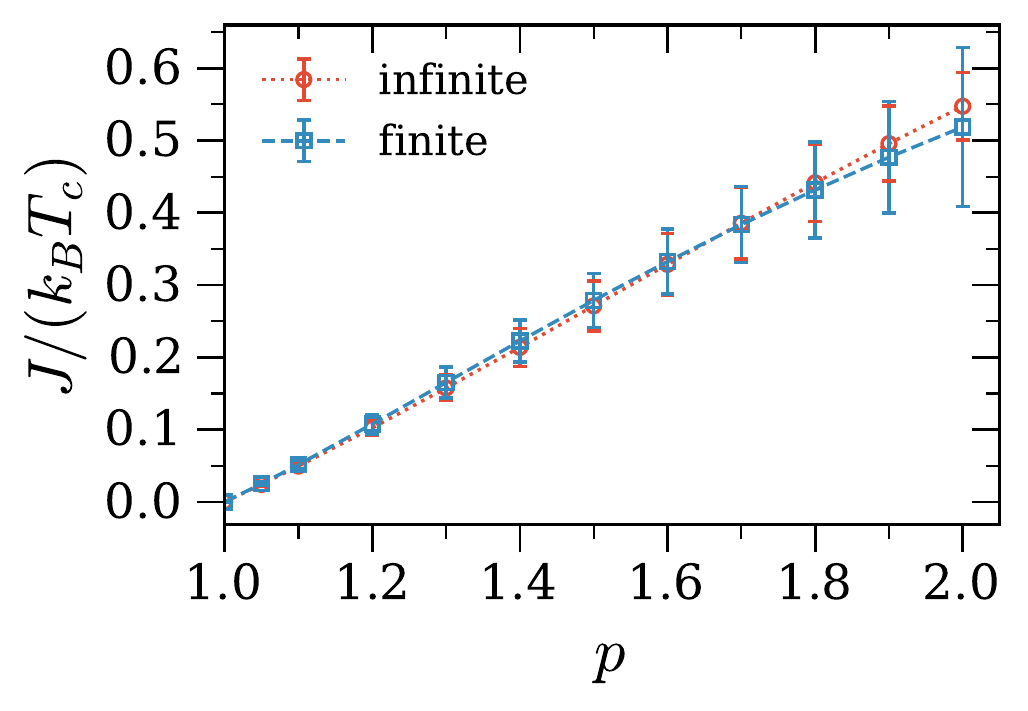}
	\caption{Extrapolated inverse critical temperature $J/(k_BT_c)$ as a function of $p$ in infinite and finite cases.}%
	\label{Fig:Tc_infinite-finiteAllEigs}
\end{figure}

Fig.~\ref{Fig:Tc_infinite-finiteAllEigs} shows the extrapolated inverse critical temperatures as functions of $p$ for both infinite and finite chains. We can see that the extrapolated temperatures are very close among them for values of $p \in \{1,1.05,1.1,1.2,1.3,1.4,$ $1.5,1.6,1.7\}$, while for $p\in \{1.8,1.9,2.0\}$, the difference is larger but smaller than a relative difference of 6\%. The critical temperature behavior could differ in both regions of $p$ due to the different types of phase transition. Table~\ref{Tab:Hurwitz_Tc} summarizes the extrapolated critical temperature given by Eq. (\ref{HurwitzExtrapolation}). In particular, for $p=2$, $J/(k_BT_c)=0.55 (6)$ and $J/(k_BT_c)=0.5 (1)$ for infinite and finite chains respectively.

\begin{figure}[h!]
	\hspace{-1.2cm}
	\includegraphics[scale=0.85]{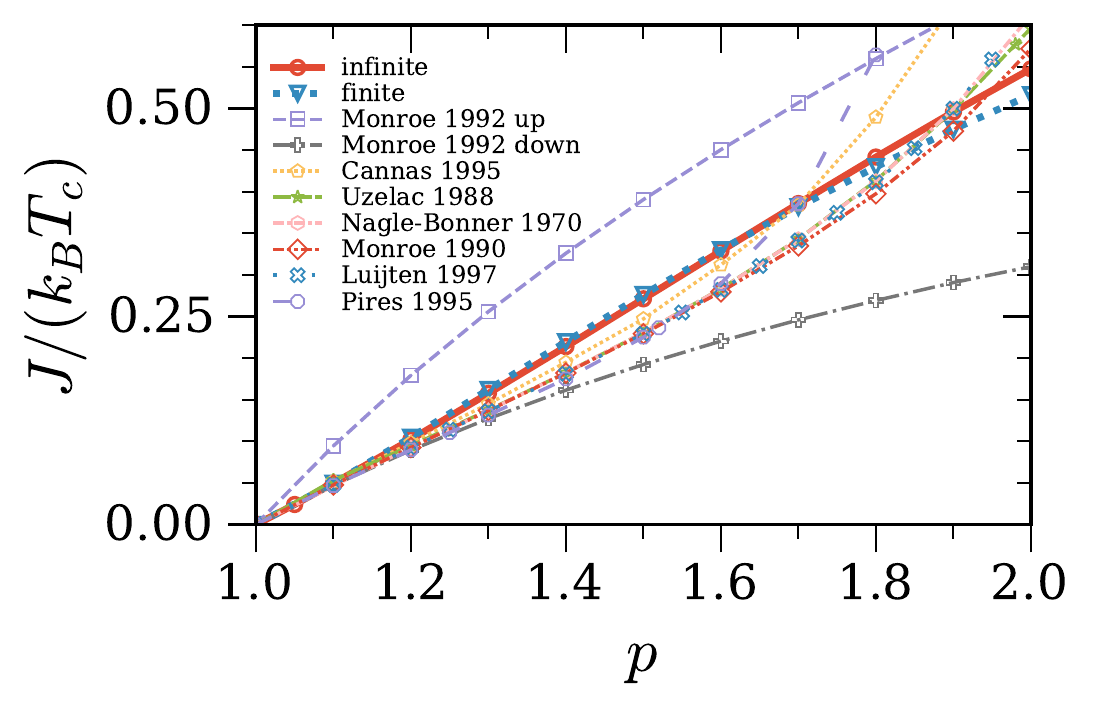}
	\caption{Extrapolated inverse critical temperatures  $J/(k_BT_c)$ as functions of $p$ for several approximations.}%
	\label{Fig:all_extrapol_inv_Tc_vs_p}
\end{figure}

To compare, in Fig.~\ref{Fig:all_extrapol_inv_Tc_vs_p}, we plot our extrapolated inverse critical temperature values together with those extrapolations given by Uzelac~\cite{Uzelac-1988}, Cannas~\cite{Cannas-1995}, Nagle and Bonner~\cite{Nagle-Bonner-1969}, Monroe~\cite{Monroe-1990,Monroe-1992}, Pires~\cite{Pires-1996}, and Luijten~\cite{Luijten-1997,Luijten-2001}. 
Our extrapolated inverse critical temperatures fall within the upper and lower bounds $(0.313,0.658)$ reported by Monroe~\cite{Monroe1994} showing a better coincidence with the plotted approximations for $p$ close to 1 than for the $p$ values near 2.

\section{Conclusions}

We have given a generalized expression of transfer matrix for a ferromagnetic spin chain of arbitrary length, including infinite length, with an exchange interaction of any finite range. We have used this transfer matrix to calculate the partition function, the Helmholtz free energy, and the specific heat for finite spin chains whose interaction ranges are equal to the variable spin chain sizes reaching $n_v = 12$, and infinite spin chains whose interaction ranges were increased until $n_v = 30$. In both cases, for each $n_v$ value, we numerically calculate the points ($1/n_v$, $J/(k_BT_{\max})$), which we interpolate and extrapolate to $n_v \to \infty$, using a novel inter-extrapolation function, $p$ and $n_v$ dependent. Both extrapolations lead to the same critical temperature within their uncertainties through two different concurrent curves for each $p$, from where we observe that the convergence rate is higher by increasing the size as well as interaction range of the finite chains than by increasing the interaction range in the infinite chains. The difference between both mentioned convergence speeds decreases as $p$ goes to 1, where the extrapolated curves are indistinguishable. The decrease of the difference in the convergence speeds, as functions of $p$, could be a signal of the phase transition type that the chains exhibit. This behavior has motivated a study in progress of the phase transition types as a function of $p$ present in the system by analyzing its correlation function properties. Finally, extrapolated critical temperatures as functions of $p$ fall within Monroe's upper and lower bounds, showing a better coincidence with many existing approximations for $p$ close to 1 than for the $p$ values near 2. This observation confirms that we have provided a direct and reliable alternative $T_c$ of the one-dimensional Ising model with long-range interaction.

\section*{Acknowledgments}
We acknowledge partial support from grants No. IN-110319 (PAPIIT-DGAPA-UNAM). We thank the Coordinación de Supercómputo, UNAM, for computing resources and technical assistance. J.G.M.-H. thanks the Ph.D. scholarship and O.A.R.-L. thanks the research assistant scholarship both from the Consejo Nacional de Ciencia y Tecnología (CONACyT), México. Also, we are grateful to the anonymous referees for their enlightening comments.


\bibliographystyle{naturemag}

\bibliography{bib_physicaA}



\appendix

\onecolumn

\section{Numerically calculated data}\label{Appendix:Tables}
%

%
\begin{table}[!htp]\centering
	\caption{Temperatures $k_BT_{\text{max}}/J$ for several values of $n_v$ and $p$, in infinite chains.}\label{tab:all_Tc}
	\scriptsize 		
\begin{adjustbox}{width=\columnwidth,center}		
	\begin{tabular}{c c c c c c c c c c c c c c}\specialrule{.2em}{.1em}{.1em} 
	\backslashbox[0.01cm]{\kern-0.4em $n_v$ \kern-2em}{\kern-1em $p$} &$1.0$ &$1.05$ &$1.1$ &$1.2$ &$1.3$ &$1.4$ &$1.5$ &$1.6$ &$1.7$ &$1.8$ &$1.9$ &$2.0$ \\ \hline\hline
		1 &0.834 (4) &0.83355 (1) &0.834 (4) &0.834 (4) &0.834 (4) &0.834 (4) &0.834 (4) &0.834 (4) &0.834 (4) &0.834 (4) &0.834 (4) &0.834 (4) \\
		2 &1.509 (4) &1.48893 (1) &1.470 (4) &1.428 (4) &1.395 (4) &1.361 (4) &1.330 (4) &1.298 (4) &1.271 (4) &1.245 (4) &1.219 (4) &1.197 (4) \\
		3 &1.979 (4) &1.93520 (1) &1.893 (4) &1.830 (4) &1.741 (4) &1.675 (4) &1.614 (4) &1.556 (4) &1.503 (4) &1.455 (4) &1.410 (4) &1.365 (4) \\
		4 &2.349 (4) &2.28281 (1) &2.218 (4) &2.115 (4) &1.995 (4) &1.893 (4) &1.801 (4) &1.723 (4) &1.649 (4) &1.584 (4) &1.523 (4) &1.469 (4) \\
		5 &2.661 (4) &2.57031 (2) &2.485 (4) &2.326 (4) &2.190 (4) &2.061 (4) &1.950 (4) &1.846 (4) &1.755 (4) &1.670 (4) &1.597 (4) &1.530 (4) \\
		\\
		6 &2.932 (4) &2.81943 (2) &2.714 (4) &2.520 (4) &2.354 (4) &2.197 (4) &2.055 (4) &1.940 (4) &1.832 (4) &1.740 (4) &1.651 (4) &1.576 (4) \\
		7 &3.174 (4) &3.03955 (2) &2.914 (4) &2.685 (4) &2.490 (4) &2.310 (4) &2.160 (4) &2.018 (4) &1.895 (4) &1.785 (4) &1.692 (4) &1.606 (4) \\
		8 &3.390 (4) &3.23678 (3) &3.093 (2) &2.832 (2) &2.605 (2) &2.407 (2) &2.233 (2) &2.081 (2) &1.947 (2) &1.829 (2) &1.726 (2) &1.634 (2) \\
		9 &3.589 (4) &3.41533 (3) &3.254 (2) &2.963 (2) &2.712 (2) &2.492 (2) &2.301 (2) &2.135 (2) &1.990 (2) &1.863 (2) &1.752 (2) &1.654 (2) \\
		10 &3.773 (4) &3.58004 (3) &3.402 (2) &3.082 (2) &2.806 (2) &2.567 (2) &2.361 (2) &2.183 (2) &2.026 (2) &1.892 (2) &1.774 (2) &1.671 (2) \\
		\\
		11 &3.943 (4) &3.73176 (4) &3.537 (2) &3.190 (2) &2.892 (2) &2.636 (2) &2.415 (2) &2.224 (2) &2.059 (2) &1.916 (2) &1.792 (2) &1.684 (2) \\
		12 &4.101 (4) &3.87378 (4) &3.663 (2) &3.289 (2) &2.970 (2) &2.698 (2) &2.464 (2) &2.262 (2) &2.087 (2) &1.937 (2) &1.808 (2) &1.695 (2) \\
		13 &4.250 (4) &4.00561 (5) &3.780 (2) &3.382 (2) &3.043 (2) &2.754 (2) &2.506 (2) &2.295 (2) &2.113 (2) &1.957 (2) &1.821 (2) &1.705 (2) \\
		14 &4.390 (4) &4.12957 (5) &3.890 (2) &3.468 (2) &3.110 (2) &2.806 (2) &2.547 (2) &2.326 (2) &2.135 (2) &1.973 (2) &1.833 (2) &1.714 (2) \\
		15 &4.524 (4) &4.24702 (5) &3.994 (2) &3.548 (2) &3.173 (2) &2.854 (2) &2.584 (2) &2.353 (2) &2.156 (2) &1.988 (2) &1.844 (2) &1.720 (2) \\
		\\
		16 &4.649 (4) &4.3581 (1) &4.092 (4) &3.624 (4) &3.231 (4) &2.899 (4) &2.618 (4) &2.379 (4) &2.175 (4) &2.002 (4) &1.853 (4) &1.727 (4) \\
		17 &4.770 (4) &4.4631 (1) &4.184 (4) &3.696 (4) &3.286 (4) &2.941 (4) &2.649 (4) &2.402 (4) &2.193 (4) &2.014 (4) &1.862 (4) &1.733 (4) \\
		18 &4.884 (4) &4.5632 (1) &4.272 (4) &3.763 (4) &3.337 (4) &2.980 (4) &2.679 (4) &2.424 (4) &2.209 (4) &2.025 (4) &1.869 (4) &1.738 (4) \\
		19 &4.993 (4) &4.6589 (1) &4.356 (4) &3.827 (4) &3.386 (4) &3.017 (4) &2.707 (4) &2.445 (4) &2.223 (4) &2.035 (4) &1.877 (4) &1.742 (4) \\
		20 &5.098 (4) &4.7503 (1) &4.435 (4) &3.887 (4) &3.432 (4) &3.051 (4) &2.732 (4) &2.463 (4) &2.237 (4) &2.045 (4) &1.883 (4) &1.745 (4) \\
		\\
		21 &5.198 (8) &4.8377 (3) &4.511 (8) &3.946 (8) &3.476 (8) &3.085 (8) &2.757 (8) &2.482 (8) &2.250 (8) &2.054 (8) &1.889 (8) &1.750 (8) \\
		22 &5.294 (8) &4.9214 (3) &4.584 (8) &4.000 (8) &3.517 (8) &3.116 (8) &2.780 (8) &2.499 (8) &2.262 (8) &2.063 (8) &1.894 (8) &1.753 (8) \\
		23 &5.39 (1) &5.002 (5) &4.65 (1) &4.05 (1) &3.56 (1) &3.15 (1) &2.80 (1) &2.51 (1) &2.27 (1) &2.07 (1) &1.90 (1) &1.76 (1) \\
		24 &5.48 (1) &5.08 (1) &4.72 (1) &4.10 (1) &3.60 (1) &3.17 (1) &2.82 (1) &2.53 (1) &2.28 (1) &2.08 (1) &1.90 (1) &1.76 (1) \\
		25 &5.56 (1) &5.15 (1) &4.79 (1) &4.15 (1) &3.63 (1) &3.20 (1) &2.84 (1) &2.54 (1) &2.29 (1) &2.08 (1) &1.91 (1) &1.76 (1) \\
		\\
		26 &5.646 (6) &5.227 (6) &4.849 (6) &4.199 (6) &3.667 (6) &3.227 (6) &2.862 (6) &2.558 (6) &2.303 (6) &2.090 (6) &1.912 (6) &1.763 (6) \\
		27 &5.727 (6) &5.297 (6) &4.909 (6) &4.244 (6) &3.700 (6) &3.252 (6) &2.880 (6) &2.571 (6) &2.312 (6) &2.096 (6) &1.916 (6) &1.765 (6) \\
		28 &5.806 (6) &5.364 (6) &4.967 (6) &4.287 (6) &3.732 (6) &3.275 (6) &2.897 (6) &2.583 (6) &2.321 (6) &2.102 (6) &1.919 (6) &1.767 (6) \\
		30 &5.96 (2) &5.49 (2) &5.08 (2) &4.37 (2) &3.79 (2) &3.32 (2) &2.93 (2) &2.61 (2) &2.34 (2) &2.11 (2) &1.93 (2) &1.77 (2) \\
		\specialrule{.2em}{.1em}{.1em} 
	\end{tabular}\label{Tab:all_temperature_Cv-max}%
\end{adjustbox}
\end{table}

%
\begin{table}[H]\centering
	\caption{Temperatures $k_BT_{\text{max}}/J$ for several values of $n_v$ and $p$, in finite chains.}\label{tab:finite_all_Tc}
\scriptsize 		
\begin{adjustbox}{width=\columnwidth,center}		
	\begin{tabular}{c c c c c c c c c c c c c c}\specialrule{.2em}{.1em}{.1em} 
		\backslashbox[0.01cm]{\kern-0.4em $n_v$ \kern-2em}{\kern-1em $p$} &$1.0$ &$1.05$ &$1.1$ &$1.2$ &$1.3$ &$1.4$ &$1.5$ &$1.6$ &$1.7$ &$1.8$ &$1.9$ &$2.0$ \\ \hline\hline
		2 &1.668 (3) &1.668 (3) &1.668 (3) &1.668 (3) &1.668 (3) &1.668 (3) &1.668 (3) &1.668 (3) &1.668 (3) &1.668 (3) &1.668 (3) &1.668 (3) \\
		3 &2.109 (3) &2.085 (3) &2.061 (3) &2.018 (3) &1.977 (3) &1.938 (3) &1.904 (3) &1.870 (3) &1.839 (3) &1.810 (3) &1.783 (3) &1.757 (3) \\
		4 &2.458 (3) &2.410 (3) &2.364 (3) &2.278 (3) &2.200 (3) &2.128 (3) &2.063 (3) &2.001 (3) &1.945 (3) &1.894 (3) &1.846 (3) &1.801 (3) \\
		5 &2.748 (3) &2.678 (3) &2.612 (3) &2.489 (3) &2.376 (3) &2.273 (3) &2.181 (3) &2.097 (3) &2.020 (3) &1.950 (3) &1.885 (3) &1.827 (3) \\
		\\
		6 &3.001 (3) &2.909 (3) &2.824 (3) &2.663 (3) &2.521 (3) &2.391 (3) &2.275 (3) &2.170 (3) &2.076 (3) &1.991 (3) &1.913 (3) &1.842 (3) \\
		7 &3.225 (3) &3.113 (3) &3.008 (3) &2.815 (3) &2.642 (3) &2.490 (3) &2.353 (3) &2.232 (3) &2.121 (3) &2.023 (3) &1.934 (3) &1.854 (3) \\
		8 &3.427 (3) &3.296 (3) &3.172 (3) &2.948 (3) &2.750 (3) &2.576 (3) &2.420 (3) &2.282 (3) &2.159 (3) &2.049 (3) &1.950 (3) &1.861 (3) \\
		9 &3.610 (3) &3.460 (3) &3.319 (3) &3.066 (3) &2.846 (3) &2.649 (3) &2.478 (3) &2.324 (3) &2.189 (3) &2.070 (3) &1.961 (3) &1.866 (3) \\
		10 &3.778 (3) &3.612 (3) &3.455 (3) &3.174 (3) &2.930 (3) &2.716 (3) &2.528 (3) &2.364 (3) &2.217 (3) &2.087 (3) &1.972 (3) &1.870 (3) \\
		\\
		11 &3.935 (8) &3.751 (8) &3.580 (8) &3.273 (8) &3.008 (8) &2.776 (8) &2.573 (8) &2.396 (8) &2.240 (8) &2.103 (8) &1.980 (8) &1.872 (8) \\
		12 &4.081 (8) &3.880 (8) &3.695 (8) &3.363 (8) &3.078 (8) &2.830 (8) &2.615 (8) &2.426 (8) &2.261 (8) &2.116 (8) &1.988 (8) &1.874 (8) \\
		\specialrule{.2em}{.1em}{.1em}
	\end{tabular}
\end{adjustbox}
\end{table}

%
\begin{table}[!htp]\centering
	\caption{Free parameters $a_1$ and $a_2$ adjusted with the expression (\ref{HurwitzModel}).}\label{Tab:Hurwitz_params}
	\scriptsize
	\begin{tabular}{cccccc
		}\specialrule{.2em}{.1em}{.1em}
		&\multicolumn{2}{c}{infinite } &\multicolumn{2}{c}{finite} \\\cmidrule{2-5}
		$p$ &$a_1$ &$a_2$ &$a_1$ &$a_2$ \\ \hline \hline
		1 &0.47 (3) &2.4 (3) &0.48 (5) &2.4 (4) \\
		1.05 &0.47 (3) &2.4 (3) &0.49 (5) &2.2 (3) \\
		1.1 &0.47 (3) &2.3 (3) &0.49 (5) &2.1 (3) \\
		1.2 &0.46 (3) &2.2 (2) &0.50 (5) &1.9 (2) \\
		1.3 &0.45 (3) &2.1 (2) &0.51 (5) &1.7 (2) \\
		1.4 &0.45 (3) &2.0 (2) &0.53 (4) &1.5 (1) \\
		\\
		1.5 &0.45 (3) &1.9 (2) &0.57 (4) &1.4 (1) \\
		1.6 &0.45 (3) &1.8 (1) &0.62 (5) &1.27 (8) \\
		1.7 &0.46 (3) &1.7 (1) &0.69 (5) &1.13 (7) \\
		1.8 &0.49 (3) &1.60 (9) &0.82 (6) &0.99 (6) \\
		1.9 &0.52 (3) &1.49 (7) &1.08 (9) &0.83 (4) \\
		\\
		2 &0.57 (2) &1.38 (5) &1.7 (2) &0.64 (4) \\ \specialrule{.2em}{.1em}{.1em}
	\end{tabular}
\end{table}

%
\begin{table}[!htp]\centering
	\caption{Extrapolated critical temperatures $k_BT_c/J$ from expression (\ref{HurwitzExtrapolation}).}\label{Tab:Hurwitz_Tc}
	\scriptsize
	\begin{tabular}{cccc}\specialrule{.2em}{.1em}{.1em}
		$p$ & infinite chain & finite chain \\
		\hline \hline
		1 &$\infty$ &$\infty$ \\
		1.05 &41 (3) &40 (4) \\
		1.1 &20 (2) &19 (2) \\
		1.2 &9 (1) &9 (1) \\
		1.3 &6.3 (7) &6.0 (8) \\
		1.4 &4.7 (6) &4.5 (6) \\
		\\
		1.5 &3.7 (5) &3.6 (5) \\
		1.6 &3.0 (4) &3.0 (4) \\
		1.7 &2.6 (3) &2.6 (4) \\
		1.8 &2.3 (3) &2.3 (3) \\
		1.9 &2.0 (2) &2.1 (3) \\
		\\
		2 &1.8 (2) &1.9 (4) \\
		\specialrule{.2em}{.1em}{.1em}
	\end{tabular}
\end{table}

%
%

\section{Specification of our calculation code and extrapolation comparisons}\label{Appendix:Program specification}

We made a Python program to do all the numerical calculations. Our code performs the following steps: 
\begin{enumerate}
	\item We fill the non-zero transfer matrix elements using the spins interactions, temperature, and magnetic field extracted from a configuration file, and we accelerate the process using the \href{http://www.latex-tutorial.com}{Numba package}.
	\item We use the \href{https://github.com/scipy/scipy}{Scipy package} to calculate eigenvalues. For infinite chains, we use the \textit{scipy.sparse.linalg.eigs} method for a sparse matrix to obtain the maximum eigenvalue. The eigensolver for this method is ARPACK. For finite chains, we use the \textit{scipy.linalg.eigvals} method for the dense matrix to obtain all eigenvalues. We note that the transfer matrix for $h=0$ is centrosymmetric, a feature that can simplify the eigenvalues calculus from a matrix of shape $2^{n_v} \times 2^{n_v}$ to a matrix of shape $2^{n_v-1}\times 2^{n_v-1}$. This simplification goes from the fact that transfer matrix $W$ is orthogonally similar to
	\begin{equation}
		\begin{pmatrix}
			A + JC & 0 \\
			0 & A - JC
		\end{pmatrix}
	\end{equation}
	where
	\begin{equation}
		W = \begin{pmatrix}
			A & B \\
			C & D
		\end{pmatrix},
	\end{equation} 
	with $A, B, C$ and $D$ are matrix blocks with half files and columns of $W$, see Ref. \cite{Weaver-1985}.
	\item We compute the expression (\ref{HelmholtzFreeEnergy}) for finite chains and (\ref{HelmholtzFreeEnergyTL}) for infinite chains.
	\item We repeat steps 1-3 for many temperature samples, we interpolate the free Helmholtz values with \textit{scipy.interpolate} method to get a curve, and then we compute specific heat with expression (\ref{SpecificHeat}).
	\item We extract the temperature at the maximum of the specific heat curve obtained in step 4. 
	\item We repeat steps 1-5 for many values of $n_v$.
	\item We fit the $(1/{n_v},J/(k_BT_{\max}))$ points with a given model through a nonlinear least-squares approach using the \textit{scipy.optimize.minimize} routine. We estimate the parameter uncertainties from the same nonlinear least-squares approach~\cite{Press-NumericalRecipes-2007}.
	\item Finally, we extrapolate the adjusted function to $J/(k_BT_{c})$ making $1/n_v \rightarrow 0$, and we compute the critical temperature uncertainty as to the maximum difference between all extrapolation values with all the maxima and minima parameters combinations. 
\end{enumerate}

The hardware we used has the following specifications:
\\

\begin{center}
\begin{tabular}{ll}
	CPU(s): & 20 \\
	Model name: & Intel(R) Xeon(R) CPU E5-2618L v4 @ 2.20GHz \\
	CPU MHz: & 1200.117 \\
	RAM Memory: & 65G
\end{tabular}
\end{center}

\begin{figure}[h!]
	\centering
	\includegraphics[scale=0.9]{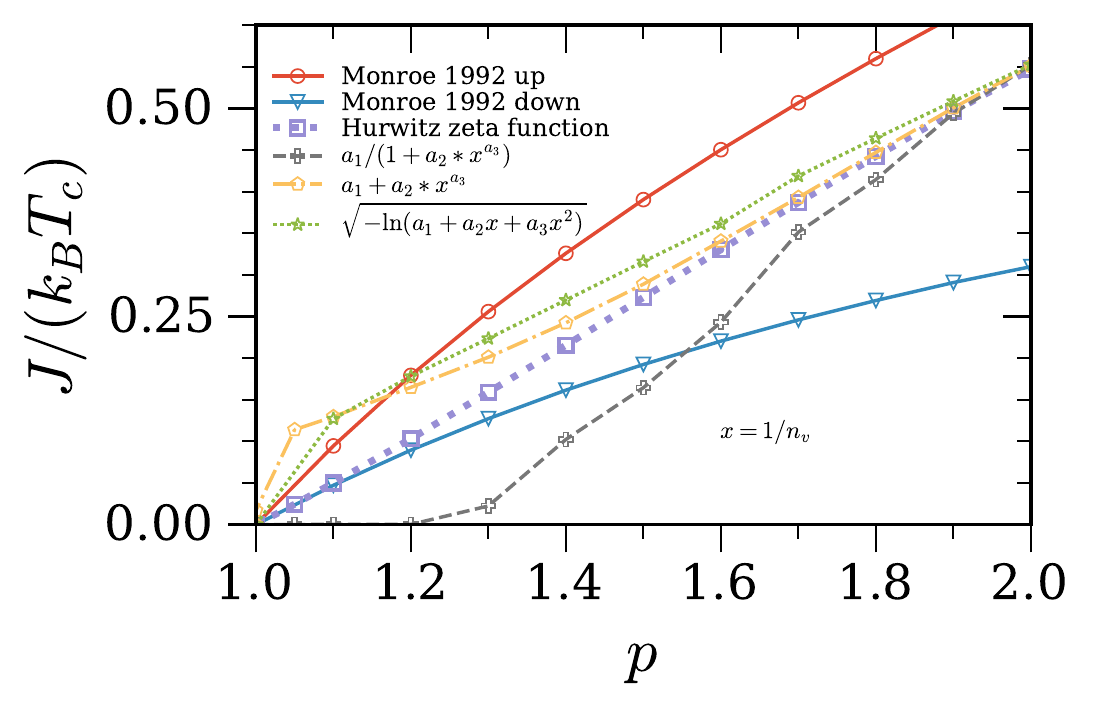}
	\caption{Extrapolated inverse critical temperatures  $J/(k_BT_c)$ as functions of $p$ for several expressions.}%
	\label{Fig:all_extrapol_models_inv_Tc_vs_p}
\end{figure}

Figure~\ref{Fig:all_extrapol_models_inv_Tc_vs_p} presents three inter-extrapolation functions, in addition to the Hurwitz zeta function, tested to interpolate the points ($1/{n_v},J/(k_BT_{\max}$) with $n_v \in \{2,3,4,\ldots,30\}$. However, the extrapolated critical temperatures with them do not fall within the Monroe's upper and lower bounds for the entire interval of $p$.

For additional details and the source code program, see Ref.
[Rodríguez-López, O. A., \& Martínez-Herrera, J. G. (2021). One-dimensional Ising model study (Version 1.1.0) [Computer software]. https://github.com/oarodriguez/isingchat].

\end{document}